# A Case for A Collaborative Query Management System

Nodira Khoussainova, Magdalena Balazinska, Wolfgang Gatterbauer,
YongChul Kwon, and Dan Suciu

Department of Computer Science and Engineering,
University of Washington, Seattle, WA, USA

{nodira, magda, gatter, yongchul, suciu}@cs.washington.edu

## ABSTRACT

Over the past 40 years, database management systems (DBMSs) have evolved to provide a sophisticated variety of data management capabilities. At the same time, tools for managing queries over the data have remained relatively primitive. One reason for this is that queries are typically issued through applications. They are thus debugged once and re-used repeatedly. This mode of interaction, however, is changing. As scientists (and others) store and share increasingly large volumes of data in data centers, they need the ability to analyze the data by issuing exploratory queries. In this paper, we argue that, in these new settings, data management systems must provide powerful query management capabilities, from query browsing to automatic query recommendations. We first discuss the requirements for a collaborative query management system. We outline an early system architecture and discuss the many research challenges associated with building such an engine.

## 1. INTRODUCTION

Modern database management systems (DBMSs) provide sophisticated features to assist users in organizing, storing, managing, and retrieving data in a database. They provide, however, only limited capabilities for managing the queries that users issue on the data. These capabilities are limited to query-by-example [28, 37], graphical tools for composing queries [5, 6], and query logging aimed at physical tuning [26, 32, 33]. Traditionally, more elaborate query management was not necessary because applications would only issue canned queries over the data (e.g., accounting or inventory management applications). These queries were developed once and used repeatedly. Emerging applications in the area of large-scale scientific data management and industrial data analysis, however, are challenging this traditional DBMS usage pattern and, as we argue, could greatly benefit from more advanced query management tools.

Scientists in areas such as biology, physics, astronomy and the geosciences collect, store, retrieve, explore and analyze vast amounts of data. Examples of large-scale scientific databases include the Sloan Digital Sky Survey (SDSS) [31], the Incorporated Research Institutions for Seismology (IRIS) [19], and soon the Large Synoptic Survey Telescope (LSST) [22]. To give an idea for the scale of these projects, the LSST is estimated to generate fifteen terabytes of raw data per night [22] for a total of five petabytes per year. Similarly, analysts and engineers in companies such as Google, Microsoft, Amazon, eBay, and Yahoo! process massive data logs collected from the large-scale services that these companies provide (e.g., clickstreams, search logs, network flow data, etc.). They analyze these logs to make informed business decisions and to improve their services. In both of these new environments, and in contrast to traditional database settings, users need to execute continuously changing queries; as the desired analysis on the data changes over time, so do the queries. Furthermore, the volume of these data sets pushes toward a usage model where the data sits in a shared data center and all users execute queries against this shared data store. Due to the increasing size of data sets, the cost of running a query has escalated, making the traditional trial-and-error method for query development too expensive. Executing queries over the whole data set is still necessary, however, because executing queries only on a small data sample can be insufficient for identifying interesting analyses.

In this paper, we argue that these new environments could greatly benefit from sophisticated query management capabilities. When users continuously develop new queries, they need support for formulating these queries. More importantly, they should leverage knowledge about queries that they or others have previously executed on the data. Such information can help both in query formulation and in deciding what queries to ask. For example, consider a scientist who decides to correlate two datasets. If the system automatically recommends previous queries correlating the same datasets but authored by another member of the research lab, the user could either quickly determine that his analysis has already been done or could leverage the existing query to author his own, especially if the old query is documented with annotations. In general, when many users submit a variety of queries over a shared database, logging, organizing, and mining these queries can produce a wealth of information. The role of a query management system is to effectively digest and present such information to users.

We propose to build a Collaborative Query Management System (CQMS) targeted at these new, large-scale, shared-data environments. A CQMS should enable users to perform simple tasks such as browse the log of all queries they submitted and document their queries by annotating them. By doing so, users will be able to quickly find, edit and re-execute past queries. Even better, the system should support sophisticated search capabilities allowing users to identify queries that operate on specific input data, have desired properties (e.g. small result set, fast execution time), or produce specific results. The system should also mine its query log and actively recommend queries to users, thus, helping them further lever-age previously performed analyses. When the data or the schema changes, the system should also monitor and evaluate the validity of the queries in its extensive collection, flagging, automatically repairing, or even removing those queries that have become obsolete. Of course, clear access control rules must be set to restrict knowledge transfer to only group members collaborating with each other.

Building such a CQMS raises several important technical challenges. First, managing a collection of queries is more akin to managing an evolving set of source code snippets rather than managing ordinary data. Queries have elaborate semantics and complex relationships with each other. The appropriate data model, query language, and features for a CQMS must thus be carefully designed to leverage the sophistication of query objects without overwhelming the user. Second, the CQMS must be efficient because it must provide hints and recommendations interactively, as a user types a new query. Finally, the CQMS should include a client with an intuitive, easy-to-use interface and effective visualization methods.

In this paper, we first explore the various query management functions that a CQMS should provide (Section 2). We sketch an initial, system architecture for a CQMS (Section 3) and discuss the many research challenges involved in building such a system (Section 4). We describe related work in Section 5 and conclude in Section 6. Overall, as the database community builds new data management systems for large-scale data repositories [1, 7, 12, 27], we argue that these systems should include advanced query management capabilities in the form of a CQMS engine.

## 2. REQUIREMENTS

In this section, we explore the different features that a Collaborative Query Management System should provide to its users. We categorize these features according to four modes of user interaction. In the default *Traditional Interaction Mode*, users simply ask queries over the data with minimal interference from the CQMS. In the *Search and Browse Interaction Mode*, users search for queries or browse through a set of queries. In the more advanced *Assisted Interaction Mode*, the CQMS interactively assists the user in composing a query. Finally, in the *Administrative Interaction Mode*, users and administrators can perform various maintenance tasks on stored queries. We now present each mode of interaction, along with the required features in more detail. We discuss resulting research challenges later in Section 4.

### 2.1 Traditional Interaction

In the *Traditional Interaction Mode*, the user submits standard queries over the data without support from the CQMS. At the same time, in the background, the CQMS logs and pre-processes these queries for further use. It is essential that the CQMS does not impose significant runtime overhead.

In addition, the CQMS should support and occasionally even request *query annotations* in this mode, especially for queries that are difficult to re-use without proper documentation (e.g. queries with more than a specified number of tables, or queries that include nesting). Users can add annotations to whole queries or query fragments. For example, a user could explain the choice for using an outer join in a query, or describe their information goal for the particular query. In general, through annotations, users can capture semantic information about their queries.

### 2.2 Search and Browse Interaction

One essential query management feature is the ability for users to search for and browse through past queries. We refer to this mode of interaction as the *Search and Browse Interaction Mode*.

**Query feature relations**

```
Queries(qid,qText)
DataSources(qid,relName)
Attributes(qid,attrName,relName)
Predicates(qid,attrName,relName,op,const)
```

**Meta-query**

```
SELECT   Q.qid, Q.qText
FROM     Queries Q, Attributes A1, Attributes A2
WHERE    Q.qid = A1.qid AND Q.qid = A2.qid
         AND A1.attrName = 'salinity'
         AND A1.relName = 'WaterSalinity'
         AND A2.attrName = 'temp'
         AND A2.relName = 'WaterTemp'
```

**Figure 1: Example meta-query for "find all queries that correlate water salinity with water temperature data".**

**Search.** A *meta-query* is a query that searches for queries. Such queries enable users to locate past queries matching specific search conditions. The resulting queries can then be learned from, re-executed, or used as a starting point to compose a new query. A CQMS should offer a range of meta-querying techniques. At minimum, it should provide substring matching and keyword search, like existing systems [29, 33]. Beyond this, we propose three other paradigms for meta-querying that a CQMS should offer.

Unlike keyword and substring search, *query-by-parse-tree* allows users to formulate conditions on the structure of the query. With this technique, the user can precisely specify conditions on the joined relations, selections, projections, nested subqueries, etc. Although powerful in searching for queries based on the query text only, this technique does not consider the query output, the query execution time, or other important information about the query as a whole.

The second technique, *query-by-data*, enables users to set conditions on the query output. The user specifies that the query output should include or exclude specific tuples. For example, the user may remember that there are some properties that distinguish Lake Washington from Lake Union. So they would request "all queries whose output includes Lake Washington but not Lake Union", and see that all matching queries specify 'temp < 18'. This problem is related to machine learning; the user specifies positive and negative training examples (tuples), and the system finds classifiers (past queries) that separate these examples. Supporting query-by-data efficiently is a challenging problem.

*Query-by-feature* involves extracting and storing specific query features in new relations. Features can be syntactic (e.g. relations in the FROM clause, predicates in the WHERE clause, attributes in the SELECT clause), metadata (e.g. author, execution time) or semantic (e.g. cardinality, samples from output). Users are able to issue SQL meta-queries that search for queries whose features match specific conditions. For example, "find me all queries that correlate water salinity with water temperature data" can be expressed as shown in Figure 1. Since such statements may be awkward to write, the CQMS could automatically generate these statements from partially written queries (at least for syntactic features). For example, if a user types: SELECT FROM WaterSalinity, WaterTemperature, the system can automatically issue the query shown in Figure 1. Due to its efficient storage in relations and flexibility in which features to capture, *query-by-features* may offer a good trade-off between expressibility and efficiency.

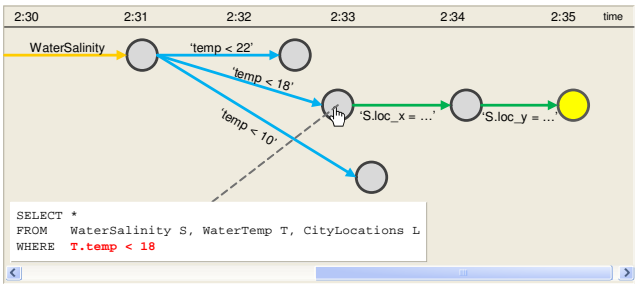

**Figure 2: Query session window of the query browser.**

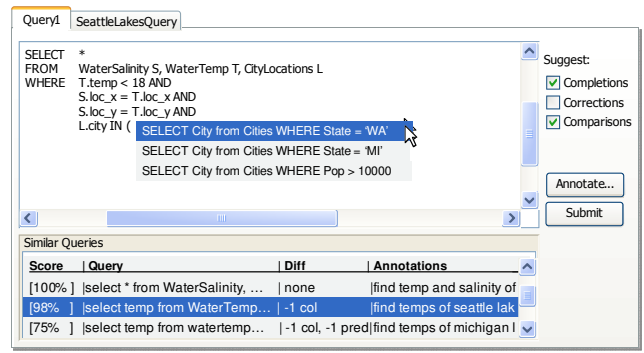

**Figure 3: Query composition in the assisted interaction mode.**

**Browse.** After finding the desired queries, the CQMS must allow the user to browse the results. Many systems that provide query logging [11, 15, 26, 32, 33] also allow the user to view the log in a table or a file. However, to make the query log suitable for browsing, the CQMS needs to present it in a comprehensible, summarized format. One possible method is to present *query sessions* instead of individual queries. A query session is a series of (often similar) queries with the same information goal in mind. Such query sessions should be automatically identified, highlighted, and visually summarized. For example, Figure 2 displays one possible visualization of a query session; each node represents a query in the session, and the edges indicate the difference between consecutive queries. In this figure, left to right, the user has added the 'WaterSalinity' relation to the FROM clause, tried different conditions on 'temp', picked 'temp < 18', and added two more predicates to the WHERE clause. Unlike a listing of the six full SQL queries, this visualization allows the user to quickly understand and navigate the query session. Subject to access control policies, users would also benefit from browsing queries submitted by other users. In this case, an effective query log visualization tool becomes even more critical to avoid overwhelming the user.

Supporting such browse and search functionality raises important research challenges related to query representation and modeling, efficient query-log processing (clustering, mining, etc.), maintenance in face of database updates, ranking (e.g., by similarity or popularity score), interface design, and query annotations. We discuss some of these challenges further in Section 4.

### 2.3 Assisted Interaction

Composing SQL queries can be a difficult task even for technically savvy users such as scientists or analysts. Furthermore, even if a user is fluent in composing queries, she can benefit from hints regarding what queries to ask. Thus, a key role of the CQMS is to assist users in query composition. For this, we define the *Assisted Interaction Mode* where the CQMS monitors the user as she types queries and provides suggestions for completing and correcting the queries. Figure 3 shows an example of assisted interaction where the system suggests several completions for the query being composed. It also shows similar queries at the bottom.

In its simplest form, *query completion* consists in providing users with possible completions for relation and attribute names as the user types a query. Although this simple capability can already be helpful, query completion can be pushed much further. In particular, the system can make context-aware suggestions. For example, assume that the most popular table to include in the FROM clause is CityLocations. However, for queries that also include WaterSalinity, the most popular is WaterTemp. Thus, if the user has already included WaterSalinity, the system should suggest WaterTemp over CityLocations. The CQMS could similarly suggest predicates in the WHERE clause of a query and even complete subclauses of the query. This capability requires efficiently mining the query log for association rules. It also raises the important challenge of when and how to present the suggestions. The CQMS must not overwhelm the user!

The second feature we envision for the Assisted Interaction Mode is automated *query correction*. Like a spell checker, while a user types a query, the CQMS suggests corrections to relation and attribute names but also changes to entire query clauses. For instance, if a predicate causes a query to return the empty set, the CQMS could suggest similar, previously issued predicates that return a non-empty set for the query. The challenge lies in defining what constitutes an error and what corrections are useful to display.

Along similar lines, a CQMS could also perform complete *query recommendations*, showing logged queries similar to those the user recently issued. Query similarity could be defined in terms of query parse trees, features, or output data. An interesting question is how to construct ranking functions that combine similarity measures together and with other desired properties (e.g. high popularity, efficient runtime, small result cardinality, etc). Overall, we posit that such functionality could be especially useful when a user first examines a new dataset and the system guides them from their rough query attempts toward similar popular queries asked by other users.

Finally, new users often suffer from a steep learning curve when trying to articulate queries. In such a case, a tutorial or step-by-step guided instructions would greatly reduce the start up cost. However, maintaining up-to-date documentation is time consuming. By analyzing the set of all queries and the evolution of query sessions, we hypothesize that a CQMS may be able to automatically produce a tutorial on the new data set or new analysis task, e.g. the system could introduce each relation and its schema by showing the user the most popular queries that include the relation.

### 2.4 Administrative Interaction

Similar to data management, query management will require a set of administrative capabilities. Users will need the ability to delete old queries, define access control rules on their queries (e.g. sharing them only with members of the same research group), etc. We call this the *User Administrative Interaction Mode*.

Furthermore, the system administrator will need to manage the query log, give preference to ranking functions, exclude irrelevant features from the similarity functions, adjust tunable parameters such as the sample size for the query-by-data approach, mark or delete obsolete queries, run offline processes to compute query sessions, query clusters, etc. We call this the *System Administrative Interaction Mode*. Similarly to automatic physical database tuning [8], the CQMS should provide automatic query maintenance when possible. For example, the CQMS could automatically flag queries that have become corrupted due to schema changes

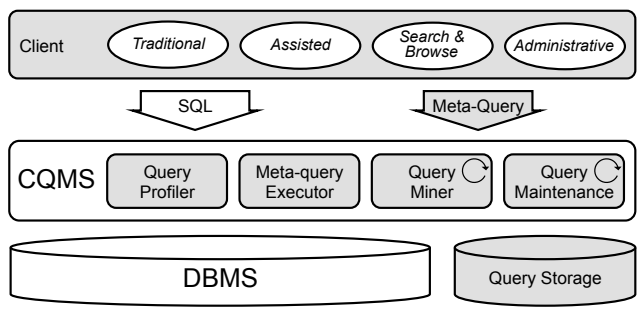

Figure 4: **QueryManager system architecture.**

## 3. SYSTEM ARCHITECTURE

In this section, we sketch a high-level design for a CQMS as shown in Figure 4. We adopt the standard client-server architecture. The CQMS client provides the four interaction modes that we outlined in Section 2 and communicates with the CQMS server through both standard SQL queries and meta-queries. The CQMS server sits on top of a standard DBMS and comprises four components.

Two of the CQMS components, the *Query Profiler* and the *Meta-Query Executor* run online: they receive inputs from the CQMS client and return results while the client waits. They must thus perform all their operations with low latency. The *Query Profiler* receives standard SQL queries as input and forwards them to the DBMS. Before doing so, however, it logs the queries in the Query Storage. Most modern DBMSs already have their own query profiling systems for performance tuning [3, 10, 29]. The CQMS Query Profiler, however, performs more sophisticated pre-processing of the queries. For example, it extracts and stores query features. It also logs statistics about the query execution and samples from its output. The *Meta-query Executor* handles all queries over the Query Storage. These queries are issued by the CQMS client during Search and Browse and Assisted Interaction modes. For Assisted Interaction modes, meta-queries may include *k*-nearest neighbors (kNN) queries. The Meta-query Executor also handles all administrative requests such as changing access control settings.

The remaining two CQMS components, the *Query Miner* and *Query Maintenance*, run in the background. The *Query Miner* analyzes the query storage. It performs tasks such as clustering queries based on similarity, association-rule mining, etc. Its goal is to extract useful information from the query log. In order to maintain all information up-to-date, it runs periodically. The *Query Maintenance* component performs automatic maintenance of the query log. Changes in the database schema or data distribution often invalidate old queries, statistics, and analysis results. The Query Maintenance component keeps the Query Storage up-to-date by flagging outdated queries and updating query statistics. The Query Maintenance can also perform physical tunings of the Query Storage to improve performance of meta-queries.

Next, we discuss the research challenges associated with these components.

## 4. RESEARCH CHALLENGES

Building a CQMS raises several important challenges. In this section, we discuss some of the challenges associated with each of the four core CQMS components and the CQMS client.

### 4.1 Query Profiler

The two fundamental challenges related to the Query Profiler are what information to capture in what format (*Data Model and Query Storage*) and how to achieve this efficiently (*Profiling Strategy*).

**Data Model and Query Storage.** A query is the primary data type in a CQMS. Queries are complex objects with elaborate structures and semantics. The data model for queries should thus effectively capture and expose these advanced query properties.

The simplest data model is to leave queries as raw text. String search is then immediately available if the underlying storage component supports it. With this model, however, it is difficult for users to express complex meta-queries that rely on query structure. This model also limits or at least complicates the query mining process.

At the other extreme, queries could be represented and stored as canonicalized parse trees using, for example, XML. The Meta-query Executor could then be an XQuery engine, thus allowing users to formulate arbitrarily complex XQuery meta-queries. The Query Miner would also benefit from this information-rich representation. However, this model leads to complex meta-queries.

An alternate data model is to represent and store queries using both their raw text and a set of pre-defined features (e.g., names of joined relations, selection and group-by predicates, and projected columns). The extracted features could be shredded and stored in a set of relations. In contrast to using XML, this model simplifies the Query Storage and Meta-Query Executor components and offers opportunities for even more efficient indexing and compression. The above features capture only syntactic query properties. However, runtime query properties can also serve for query management. Several common runtime features, including result cardinality, execution time, and the query execution plan are already incorporated in existing query profilers. In a CQMS, we envision that the system also captures the query result. This semantic query feature captures information about the intent of the query. It enables comparing queries as black-boxes.

Finally, in addition to modeling a query, a CQMS must also model *query sessions* as discussed in Section 2. A query log for a user then takes the form of a tree of query sessions where vertices correspond to individual queries, but queries are labeled with the unique ID of the session they belong to. Edges represent relations between queries as shown in Figure 2. Examples of relations between queries include temporal relations, modification relations and investigation relations (where the latter query investigates why certain tuples are included in the first query's output). The query log can be stored as a standard normalized edge relation, i.e. a pair of query identifiers and an edge type.

**Profiling Strategy.** The Query Profiler should not hinder ordinary data processing. This raises several system design questions.

*Profiling queries:* If queries are modeled as raw text or as XML parse trees, the profiling process is straightforward. Its overhead can be almost completely removed by profiling queries during the normal query-parse phase of the DBMS query execution (although this would tie the CQMS more closely with the underlying DBMS). Even without this optimization, the overhead still remains small. In contrast, if the profiler needs to extract a variety of query features, the overhead depends on the selected features and extraction algorithms. Most syntactic features should be easy to log, but if the system needs to do a variety of additional lookups, such as logging the differences with the previous query in a session, the overhead may quickly grow. Such additional tasks are thus best left to the Query Miner, even if this imposes some latency in when the data becomes available for meta-querying.

*Profiling query results:* Since it is usually infeasible to log the whole output of a query, we must find techniques to summarize the output succinctly. This problem is closely related to selectivity estimation [16] and standard approaches exist including building histograms or sampling. An interesting additional option is to adjust the maximum size allowed for the output summary depending on the query execution time. For example, if a query takes two hours to complete and outputs ten rows, then the system should store the whole output. However, if a query takes only two seconds and outputs two million rows, there is no need to store the output.

## 4.2 Meta-Query Executor

As discussed in Section 2, a CQMS could support many types of meta-queries (query-by-parse-tree, query-by-features, and query-by-data). A key question is *what query language to use for these meta-queries*. Are existing languages sufficient to express them? SQL is a viable option when queries are represented as raw text or as a set of features. XQuery is possible for the parse-tree model. However, in both cases, the meta-queries can quickly become complex. Users may need assistance in authoring the meta-queries themselves! Approaches for addressing this problem include either creating a simpler language for meta-queries or providing a tool in the CQMS client that will generate meta-queries from more intuitive user interactions (e.g. from a partly specified SQL query). Both approaches can lead to interesting research problems, although the latter technique is likely superior as it maintains the power of existing languages while providing good support for users.

Independent of query language and data model, we expect four major classes of meta-queries: keyword meta-queries, complex meta-queries explicitly stating conditions on query features or structure, meta-queries with conditions on query outputs, and kNN queries. The first two classes of queries should be straightforward to support once the query data model, meta-query language, and storage layer are designed. The latter two types of queries, however, are significantly more challenging to provide. The first challenge lies in the meta-query semantics, i.e. what it means for two queries or the output of two queries to be similar. The second challenge lies in meta-query performance. Meta-querying must be interactive, which can be challenging depending on the distance functions chosen for queries and their outputs.

## 4.3 Query Mining

The CQMS Query Storage can grow considerably over time. Mining this archive can benefit all non-traditional interaction modes. There are various types of query analysis possible. Due to space constraints, we discuss only two prominent ones.

*Clustering:* By clustering queries, a CQMS can not only save storage space (through better compression of similar queries) but can also provide more sophisticated query management capabilities. For example, the CQMS can better deduplicate meta-query results or at least group results into sets of similar queries. It can also provide better query recommendations and similarity searching. The idea of query clustering has been around for years [4, 17, 23]. However, previous tools were not designed to help database users but rather database administrators or query optimizers. Similarly, if the CQMS clusters entire query sessions, it can provide better services. For example, a meta-query could be restricted to return only results from similar query sessions, or query recommendations can be limited to queries from users who have similar query session patterns as the current user.

*Association Rules:* By learning association rules [2], which capture relationships between values in a database, a CQMS could provide more advanced support for query composition. For example, by mining common edit patterns, the CQMS could provide better completion or correction suggestions. Similarly, by mining common query evolution patterns and correlating them with query runtime features, a CQMS could automatically generate a tutorial for new users demonstrating common mistakes and good practices.

**Challenges in Mining.** First, the definition of similarity between queries should be reconsidered from the perspective of users and query management tasks. To be usable, the system needs to go beyond string similarity. Better options include parse tree similarity (perhaps after removing the constants from the tree), or looking at similarities between query features. Second, more advanced multi-relational mining [14, 36] and graph mining [9] could produce useful information since queries and sessions form a graph and can be spread across multiple relations. However, such advanced techniques could impose a greater runtime cost. Finally, incremental mining algorithms and, in general, incremental maintenance will likely be necessary considering the possibly rapid growth of the query log in a large-scale shared database.

## 4.4 Query Maintenance

The Query Maintenance component strives to maintain all information in the Query Storage up-to-date. In this section, we discuss challenges caused by schema evolution and data distribution changes, as well as the challenges of measuring and maintaining query quality.

Queries logged in the Query Storage operate on an underlying large database whose schema can change. Schema evolution can cause some of the stored queries to stop working. The CQMS should be able to efficiently identify affected queries and handle them appropriately. Identifying affected queries can be implemented by comparing the timestamp of a query with that of the last schema modification on any input relation. Handling potentially invalid queries is more challenging. A simple option is to drop such queries from the storage, but this approach may drop significant numbers of queries. Another option is to systematically repair the queries by applying appropriate changes, but how to perform such repairs automatically is an open question.

Significant changes in data distribution may invalidate runtime query features discussed in Section 4.1. A naïve and overly expensive solution is to rerun all queries periodically to renew their statistics. A better approach is to re-execute queries only when there is reason to believe their statistics have significantly changed. This solution may prove to be efficient if there is an accurate method for detecting such changes. The system could also update the statistics more frequently for popular or important queries.

In addition, the system must maintain a measure of the query quality because the system is most useful if it can quickly and accurately provide users with queries that are not only relevant but also of high-quality. The definition of high-quality may differ across various applications. For example, quality can be defined in terms of query efficiency, query simplicity, source tables' quality, etc. Measuring query quality is thus a difficult but important task.

## 4.5 CQMS Client

We envision the Assisted and the Search and Browse interactions to become the default interaction modes for scientists, industry analysts, and other users working closely with large-scale, shared data sets, especially if they are new to databases. The challenge for the CQMS client is to make these modes of interaction intuitive to use. For this, we need new interaction and visualization techniques, which we intend to adapt from results in software engineering and human-computer interaction (HCI) research. In particular, we envision the CQMS client to take the appearance of a modern, integrated development environment, such as Eclipse [34].

A *mixed-initiative interface* is one that offers both automated reasoning and the ability of direct manipulation to its users. The CQMS client is an example of such an interface. Therefore, HCI research on mixed-initiative interfaces [18] can guide us in designing a simple and intuitive interface for the CQMS client.

Our initial ideas for query completion and correction come from existing tools. For example, we propose to initially use drop-down menus, a standard technique from search interfaces and code editors, for auto-complete suggestions. For corrections, we borrow from spell checkers: indicate corrections by coloring the affected text and show a pop-up menu with details of the suggested correction when the user hovers over the text. Visualizing the difference between a query and its recommended queries is difficult because the differences may not be apparent in the query text. As a first attempt, we propose a separate sub-window to show the additional information such as the semantic similarity, popularity, date of last execution, etc. The novel challenge for the client interaction is the presence of possible completions and corrections at any point in the query and at different levels of granularity (e.g. the current token, the current predicate, or the current clause). Clearly, if not designed well, the query completion tool can be a painful work companion. Its design, implementation, and evaluation thus requires real HCI expertise.

Finally, an important aspect of query browsing is that a user should be able to quickly identify key differences and dependencies between queries. Such differences or dependencies can occur on various axes such as differences in data sources, selection predicates, annotations, popularity, execution time, etc. Thus the key challenge for the query browser is to effectively present all this information in a comprehensible and navigable visualization.

## 5. RELATED WORK

Today's DBMSs provide limited query management capabilities. The closest to our proposal are the DB2 Query Management Facility [11], Query Patroller [29] and EMS SQL Management Studio for Oracle [15]. These tools support graphically composing queries, logging queries, sharing query repositories, and viewing the query plan of stored queries. Query Patroller also analyzes queries before execution to ensure good performance. There are also systems that support only graphical composition [5, 6] and those that support query logging [26, 32, 33] but primarily for physical database tuning. Also related are tools that allow 'query by example', where the user sets the attributes of a sample object, which is then converted into a SQL query [28, 37]. However, none of these tools allow users to annotate queries or perform advanced searches over stored queries.

Many relational mining techniques [13] are directly applicable for query mining, especially multi-relational mining [14, 36] since queries are likely to be stored in multiple database relations. Related also is the work on association rule mining [2].

Query clustering is often used in database performance tuning, e.g. Aouiche *et al.* [4, 23] cluster queries to reduce the search space for the materialized view selection problem, and Ghosh *et al.* [17] use clustering for amortizing the cost of query optimization.

Work that may be adapted for query similarity includes the Context Distance Measure framework [21] designed for computing the distance between two objects defined across multiple relations. Also relevant is the work in Kim's thesis [20], which explores the structure of code change. In this work, the author explores ways of measuring the similarity between two programs' source codes. Kim also proposes a method for succinctly describing the difference between two source codes which could be helpful for tersely describing the differences between queries in a query session.

Assisting programming tasks has been a long issue in software engineering research. Recent approaches mine open-source software code repositories to extract reusable patterns and provide context-sensitive assistance like completion and recommendation within the development environment [24, 25, 30, 35]. The goal of such research closely matches that of a CQMS: help users write "correct" queries easily. A CQMS, in fact, aids new assistance tool development because the developer no longer has to consider assistance logic and data management. They simply use meta-queries.

## 6. CONCLUSION

New environments are emerging where large numbers of users need to develop and run complex queries over a very large, shared data repository. Examples include large scientific databases and Web-related data. These users are not SQL savvy, yet they need to perform complex analysis on the data and are further constrained by the high cost of running and testing their queries, often on a shared server cluster. We have argued for the need of a Collaborative Query Management System that logs and organizes all queries, and assists users in formulating new ones. Such a system, however, poses several research challenges, which we have discussed in the paper. Our goal is to address these challenges, build a CQMS, and test it in a scientific database environment.

## 7. ACKNOWLEDGMENTS

This work was partially supported by NSF Grants IIS-0713123, IIS-0454425, IIS-0627585 and IIS-0428168. The authors would also like to thank the reviewers for their helpful feedback.